# Sustaining IT PMOs during Cycles of Global Recession

**Parvez Mehmood Khan, M M Sufyan Beg, Musheer Ahmad**
*Department of Computer Engineering, Faculty of Engineering & Technology, Jamia Millia Islamia, New Delhi - 110025.*
E-mail: pmkhan@hotmail.com, mmsbeg@hotmail.com, musheer.cse@gmail.com
Tel: +91-986-8007-502 ; Fax: +91-112-698-1261

**Abstract**

Growth in the number of PMOs established by the industry over last decade and ever growing body of literature on PMO related research in academia is a clear indication that there is very clear interest of researchers, practitioners and industries across the globe to understand and explore value propositions of PMO. However, there is still a lack of consensus on many critical aspects of PMOs. While there are many PMOs being established, but there are also many being closed and disbanded, which is definitely a matter of concern. In industry environment, a narrow majority of PMOs are well-regarded by their organizations and are seen as contributing business value, many of the others "*are still struggling to show value for money and some are failing, causing a high mortality rate among PMOs*". This paper is the result of a study undertaken to get a deeper understanding of factors that may be causing mortality and failure of PMOs. Post Implementation Reviews of 4-failed & 3-challenged PMOs in IT-Industry were carried out with concerned Project Managers & PMO-staff, using grounded theory research method, with support from the concerned enterprise from IT-Industry.

## 1. Introduction

Recent global recession and economic slowdown has forced almost every organization to struggle with managing funds. As a result, there is organizational mandate to do more with less, and performance is more important than ever before. This is true for public sector, private sector entities, as well as for NGOs and other organizational entities across the globe. It is becoming crystal clear to all that no organization can sustain if its business critical projects are not managed effectively — on time, within budget, and delivering stated business outcomes for clients, whether they are internal customers or external clients. Consequently, more than ever before, business organizations are looking forward to establish a more centralized management structure for large groups of business critical projects, which is what practitioners refer as PMO.

A PMO is an organizational entity assigned various responsibilities related to the centralized and coordinated management of those projects under its domain. The responsibilities of a PMO can range from providing project management support functions to actually being responsible for the direct management of the projects[1]. PMO is being recognized as a phenomenon widely applied in practice with no solid theories underlying [3][4] . The proponents of PMOs are typically practitioners and consultants [3]. In [4], Hobbs.et.al launched a multi-phase research programme sponsored by PMI aiming to provide better understanding of PMOs and the dynamics surrounding them in their organizational context, and develop theory. In [5] Aubry.et.al suggest historical approach for understanding PMOs: "The study of the organizational processes that are behind the instability of PMOs provides a better approach than trying to find what is wrong with the current PMO and the search for an optimal design." They stress the importance of the organizational context as PMO is embedded into the host organization and both co-evolve.

Many studies on PMO have shown that often PMOs are temporary organizations, and are either disbanded or subject to radical transformations after several years [6]. However, Hurt.et.al in their research programme[7] disagree that this is a negative issue and form an alternative view that PMO still can bring and sustain value for organizations. Pellegrinelli.et.al[3] confirm that PMOs are "agents and subjects of change and renewal rather than stable, enduring entities", and offer the notion of creative destruction of replacement one form by another and generation of new value.

The only agreement seems to be that there is "no one size fits all PMO solution", and PMO should be carefully fitted to the needs of every particular organization. The discussion on the value of project management and the concept of "fit" between the project management implementation and organizational context is still ongoing [8],[9],[10]. For example, in [9] it is discussed that strategic drivers influence what value is expected from project management, and to maximize the value resulting from projects project management system should be adapted to the strategic positioning of the particular organization.

Previous research provides contradictory and inconclusive evidence on the value relevance of PMO, giving both positive and negative results. Thus, present study is an attempt to take a deep dive into negative results area and using comprehensive feedback from failed PMO initiatives to shed more lights on the issue of PMO sustenance. The objective of this research is to have a better understanding of key determinants that contribute significantly to establishing a sustainable PMO in the current scenario of global recession and cost-cuttings.

## 2. Project Management Office (PMO)

A PMO is an organizational entity assigned various responsibilities related to the centralized and coordinated management of those projects under its domain. The responsibilities of a PMO can range from providing project management support functions to actually being responsible for the direct management of the projects. The projects supported or administered by the PMO may not be related, other than being managed together. The specific form, function and structure of a PMO is tailored to the needs of the organization that it supports. In some organizations, a PMO may even be delegated the authority to act as an integral stakeholder and a key decision maker during the initiation of each project to recommend actions as required to keep business objectives of the project consistent. Not only this, but also PMO may be involved in selection, management and deployment of shared or dedicated resources on the projects.

There is plethora of literature available on PMO, including numerous models and concepts documented on organizational structure and project & programme governance framework. However, there is a very clear gap in the literature, with respect to the exact linkage or relationship between the PMO functions and project success criteria. A majority of the pre-existing literature does not appropriately explore the role of the PMO in aligning projects with the business strategy. The question that then arises is that, 'what influence does PMO have on the key performance criteria / knowledge areas[1] of project management such as cost, time, quality, resources, procurement, planning, risks and communications?' the ensuing research investigates this area by exploring the role of PMO in technical organisations and examining its impact on IT project performance. Some researcher have investigated the role of PMO in aligning IT projects activities to the organisation's strategic objectives with success.

## 2.1. Typical functions of a good PMO

In recent times of financial crisis, businesses are having greater expectations from their PMOs than ever before. Consequently, PMOs are expected not only to provides standardization, as the foundation, but to offer more business value to the sponsoring organization, like:

- ❖ Benefits Tracking;
- ❖ Expert Work Planning, Estimating & Scheduling;
- ❖ Coordinating Resource Management;
- ❖ Structured Progress Tracking and Forecasting;
- ❖ Robust Scope Management and Integrated Change Control;
- ❖ Focus on Budget Efficiency;
- ❖ Stakeholder/Communication Oversight;
- ❖ Industrialized Quality Management;
- ❖ Value-adding Risk and Issue Management;
- ❖ Comprehensive Knowledge/Records Management; and
- ❖ Fully-Integrated Project Delivery Framework/Processes.

## 2.2. PMO Maturity Levels

PMO Maturity levels indicate the **effectiveness of project execution support**. A brief explanation of different maturity levels of PMOs is articulated below:

**2.2.1** _Level 0: Absent_ - No identifiable PMO operating. No influence on project success.

**2.2.2** _Level 1: Immature/Initial_ – A named entity (identified as a PMO) is operating but, in general, processes are inconsistent, project data isn't collected/used and reporting is broadly qualitative. Has little influence on project success.

**2.2.3** _Level 2: Established/Repeatable_ – A recognised PMO is operating but is in need of improvement. Some processes have consistency, some project data is collected but little of it is analyzed and reporting is still primarily commentary based. Plays a limited role in project success.

**2.2.4** _Level 3: Grown-up/Defined Standard Processes_ – A solid PMO which experiences more successes than failures. Most processes have consistency, most key project data is collected but only basic analytics are done and reporting has introduced some metrics. Plays a role in some project successes (and failures).

**2.2.5** _Level 4: Mature/Managed_ – A very successful PMO which has good sponsorship. All core processes are consistent, all key project data is
collected, solid analytics are undertaken and reporting is primarily data-driven. Plays an important role in the success of the project
environment.

**2.2.6** _Level 5: Best in Class/Optimised_ – A world-class PMO which has complete sponsorship. All core processes are consistent and continuously improved, all key project data is collected, analytics are comprehensive and reporting is completely data-driven. Plays a critical role in the success of the project environment.

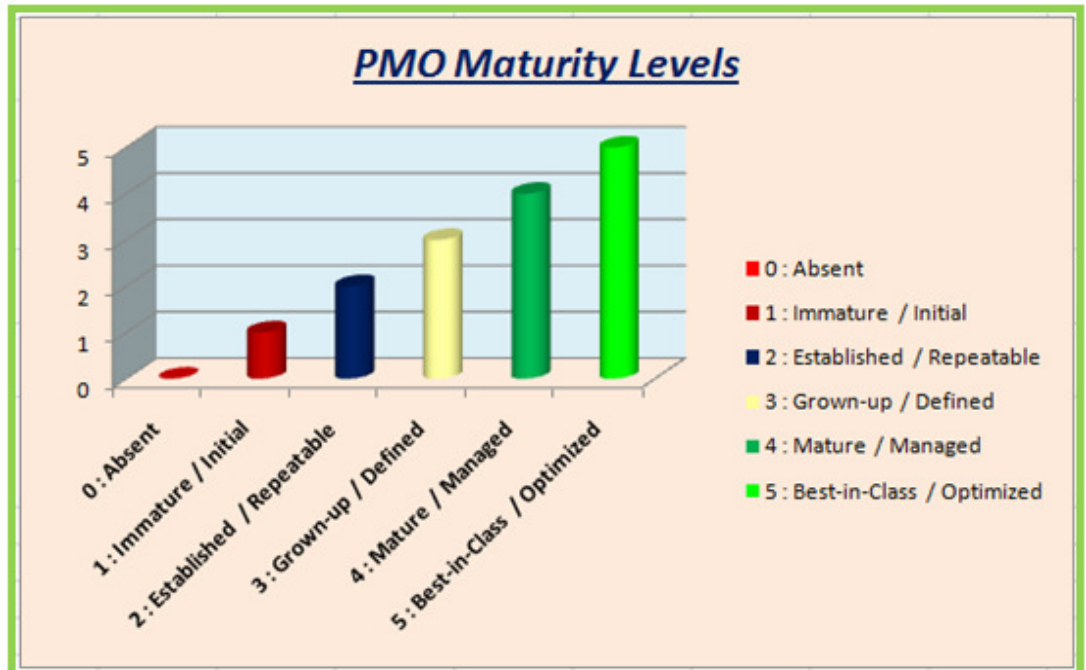

**Figure-1**: *PMO Maturity Levels in IT-Industry*

There is a common assumption that the deployment of a PMO is the panacea to project ills. Unfortunately, that is not true. The key driver in organizational performance improvements has been found by the practitioners to come from the maturity level of the PMO not just the deployment of a named PMO function. It is therefore, critical for organizations who have already setup PMOs to strive to improve the maturity level of the PMO, in order to get best return on their PMO investments.

**2.3. Industry Examples & views on PMOs**

A survey of 60 senior executives conducted by ESI International has revealed that many PMOs are being challenged on a number of fronts, including:
- ❖ The PMO was often seen as an extension of administrative support, rather than a professional body with value-add skills.
- ❖ Budget cuts necessitated cost justification, a difficulty for the usually non-revenue producing PMO.
- ❖ The PMO size and organizational setup were viewed as counter to the time constraints under which project managers operate.
- ❖ There was a lack of understanding of the business benefits of the PMO, especially among executive management.

Even in these challenging times, the PMOs with higher maturity are sustainable and doing pretty well. For example, at Siemens Product Lifecycle Management (PLM) Software, their PMO for the Europe, Middle East and Africa (EMEA) region continues to bring value both to the organization and their customers. The PMO in question oversees 100 project managers, and two hundred projects with a revenue flow of $350m per year (to give an indication of scale), as articulated by their vibrant and dynamic PMO Director Mr. Peter Taylor, in his famous "*The PMO – A Three Year Journey*" articulating details of how the PMO within PLM Software continued to be deemed "*fit for purpose*". As emphasized by Peter, the PMO at PLM Software was evolved to be seen as supportive of the organization as a whole and not grown to a size that outweighs our business benefit. Consequently, it was too valuable to lose and but equally important, it was not too expensive to keep.

Another survey from project management firm PM Solutions found that PMOs are becoming more influential and entrenched than ever. The "*State of the PMO 2012*" [11] survey reached out to more than 500 project managers across a number of industries and saw that, in general, PMOs are playing a significantly larger role in strategic functions, putting greater visibility (and greater pressures) on their success. PM Solutions noted that over the last 12 years the number of organizations with a PMO grew by 40 percent – from 47 percent in 2000 to 87 percent in 2012.

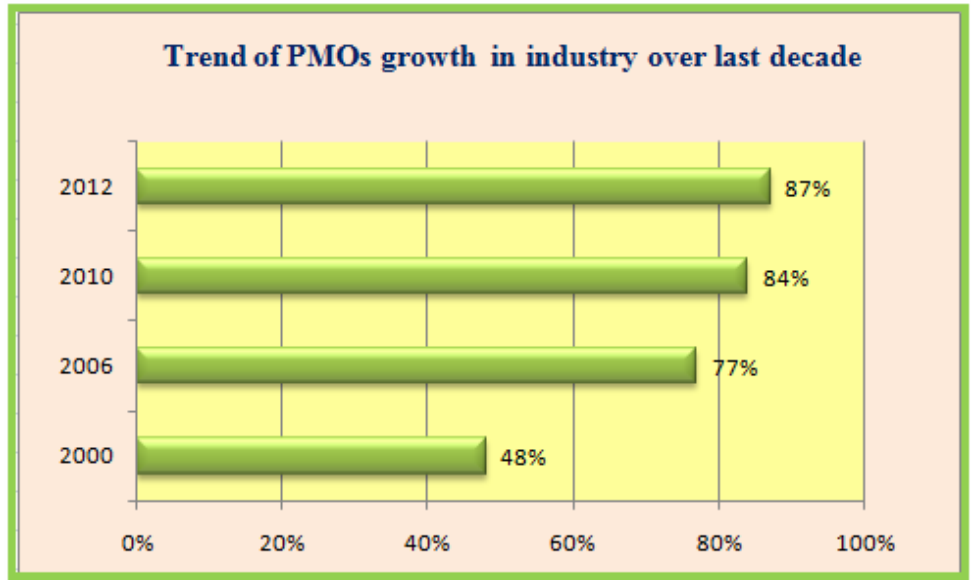

**Figure 2**: *Increasing trend of PMOs adoption by industry*

As a matter of fact, there is a steady growth in PMOs from 48% in 2000, 77% in 2006, 84% in 2010 and 87% now in 2012. This fact is depicted in Figure 7.2. The primary purpose of the PM-solutions industry research & study of "State of the PMO 2012" was to gain a clearer understanding of current PMO operations such as size, functions, staffing, challenges, performance and capability.

## 3. Research Background and Context

Globalization and the internationalization of the markets in IT-industry have increased competitive pressures on business enterprises. Recent global recession has only worsened the problem and impacted the US and Europe markets and indirectly created huge pressures on Indian IT-companies, because most of the customers and revenues for IT-majors were historically linked to US and Europe markets. These pressures have led companies to engage in projects that are not only critical to their performance, but also vital for survival of the enterprise. Majority of the companies continually strive to produce better results by undertaking strategic projects. Distressingly, industry trends have found that majority of the projects exceed budget, are completed past scheduled deadlines and do not meet original business objectives. One solution to this problem (that had been previously slow to gain popularity in SMBs and start-up organizations) is the implementation of a Project Management Office (PMO) in IT-Industry.

Software Project Management as a discipline is slowly maturing even for medium size organizations, as a result of successful deployment of PMOs nowadays. However, uncertainty level about PMO's role, implementation, relevance and value for many organizations is still not clearly understood, leading to challenging the PMO initiatives and questioning the relevance of PMOs. The objective of this research is to have a better understanding of key determinants that contribute significantly to establishing a sustainable PMO in the current scenario of global recession and cost-cuttings.

## 4. Previous Research

We began this research with thorough examination of the existing literature that deals with PMO research conducted so far, in order to identify open research issues pertaining to PMO.

In the research work of Martin, Pearson & Furumo[2], an attempt has been made to empirically investigate the role of a PMO in IS projects, but insufficient sample consideration has limited the outcome. The study was more focused in projects in Academia and did not have sufficient blend of Industry projects, inclusion of which would provide better sample and more reliable empirical results for SDLC-projects which are under consideration in this research work.

In yet another research work[12], conducted at the Department of technology of the Universidad Técnica Particular de Loja, in collaboration with PMO members (who worked day to day for getting results in each projects they manage) and published by J Carrillo and others, there is a debate about Success Factors for creating a PMO aligned with the objectives and organizational strategy which aims at setting-up a functional PMO which adds value to the organization.

In another Industry oriented Research conducted by Gartner [13], it has been found that "building a PMO is a timely competitive tactic". Based on their research, Gartner published industry forecast that organizations who establish standards for project management, including a PMO with project governance, will experience half the major project cost overruns, delays and cancellations as compared to those that fail to do so.

In another research work[14] carried out with an effort to ensure compliance and consistency in and R&D environment, the Princeton Plasma Physics Laboratory (PPPL) established an office of PMO which was responsible for oversight, coordination and implementation of all project management processes. Establishing PMO had helped overcome the challenges associated with the use of legacy processes by introduction of several new processes, the PPPL Office of Project Management has established oversight and coordination of projects at PPPL. Post implementation review of PPPL PMO has shown positive results with respect to ultimate goal of doing more with less at PPPL.

In another research work[15], it is discussed that by managing project delivery through PMOs, organizations are able to improve project delivery performance, and as a result organizations' satisfaction rates are increased; hence, improving value to the business which in turn leads to improving overall organizational performance.

In another research work[16], it is discussed that improving overall performance for business organizations can be achieved through improving project delivery success rates, and minimize issues leading to failure which led for many organizations to establish PMOs.

In another research work[4], it is discussed that organizations having a PMO are likely to have better performance than organizations that do not. Over the last decade, one of the reasons that a number of organizations have created a PMO has due to the need for better managing project delivery.

In another research work[17], it is found that PMO is a critical organizational entity that, although it may differ in types and may perform different functions, should be focused on contributing to competitive advantage and adding value to an organization and its customers to achieve desired organizational performance.

In another research work[18], it is found that Software Development Projects can get value propositions if delivered under a well established PMO. It is also found that PMOs are still evolving and there is plenty of scope for further research in industry environment for establishing PMOs that are sustainable and offer the business value to the organization.

As noted above, little work has been devoted to detailed analysis of failed PMOs in IT-Industry to identify and document key lessons learnt from such failures, which can help avoid recurrence of PMO failures in future. This research paper is an attempt to address this issue, by making a detailed assessment of resources who had been associated with failed and seriously challenged PMOs from IT-Industry.

## 5. Research Methodology

This research is exploratory in nature and examines failed and seriously challenged PMOs from customer base of small and medium size IT-MNCs having their ODC(Offshore Development Center) in India. In this research, we interviewed project managers & practitioners from different IT-MNCs all of whom were part of a SPIN(Software Process Improvement Network) in India. Selection of practitioners through SPIN-forum was done because of, ease of access to experienced professionals at the same time (during weekend meetings, workshops & trainings frequently conducted by SPIN), as well as good cross section of companies and their non-business data (available in public domain) frequently shared an SPIN meets. Grounded theory research methods were adopted for this research for two reasons. First, the research was aimed at the extension of existing theory. Grounded theory is generally deemed appropriate for such efforts as it allows theory to emerge from interview transcripts and organization artifacts. Second, the methodology is reputed to help separate researcher biases from interpretation of the data.

A total of 49-practitioners (project managers and PMO executives) were interviewed over a period of 7-months. The project managers chosen for the interview were generally very experienced in their position with average experience of more than 12-years and had managed project budgets of multi-million-dollars. Four of the executives were retirees, having left the organization within the preceding couple of years.

The interviews were facilitated via open-ended questions intended to encourage individuals to describe those PMO attributes that they think are reasons for PMO failures or challenges posed to existence of PMO, during their association with PMOs in IT-Industry. The resulting pages of interview transcripts were analysed to identify recurrent themes and concepts associated with good and bad times in the business. Those themes and concepts were then tested against archived data including formal documents maintained by PMO (like PMO-Charter, Project Data & Metrics Reports, monthly PMRs, MOMs and other formal documents available) with an eye to find evidence reinforcing or challenging the themes and concepts. Preliminary conclusions were then tested via follow-up interviews with a subset of original 49-individuals. This iterative process (*interviews, concepts identification, concepts aggregation, theme analysis, testing against organizational artifacts, review with interviewees, adaption, repeat*) led to the identification of key concepts and vital lessons learnt, for PMO sustenance. These findings are articulated in the following sections.

## 6. Discussion on our Research Findings
This study complements current research and draws attention to an area in PMO research that is largely overlooked. Majority of the current research on PMO primarily focuses on defining the value of PMOs and how to quantify and measure those values for ROI justifications of PMOs. In this research, we have focused on failed and seriously challenged IT PMOs to investigate and get a better understanding of what lessons can be applied to fresh PMOs in order to make them sustainable. In this section of paper we present the analysis of our result, which are graphically represented in figure-3, figure-4, figure-5 and figure-6 respectively and also discussed below.

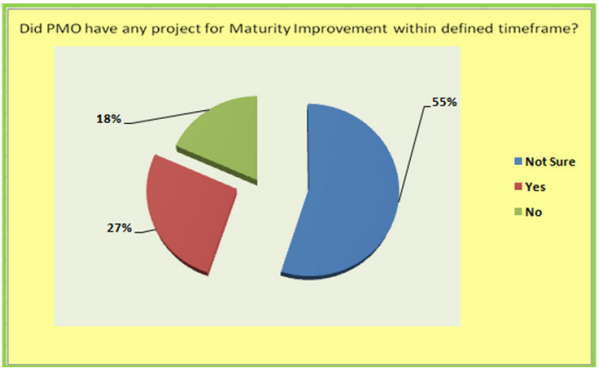

**Figure 3**: *Summary of Key Results Set #1*

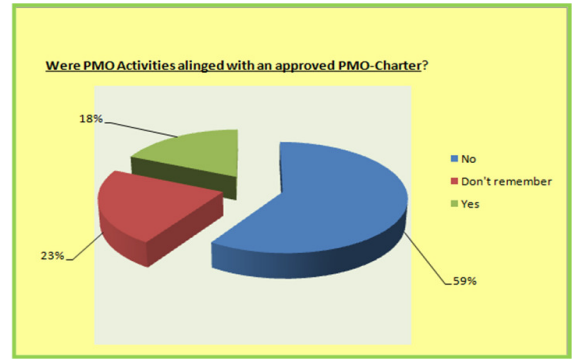

**Figure 4**: *Summary of Key Results Set #2*

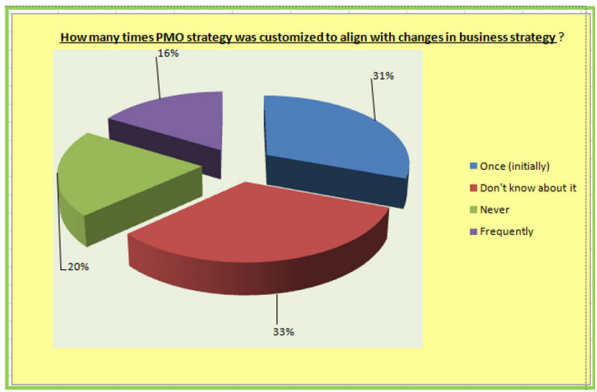

**Figure 5**: *Summary of Key Results Set #3*

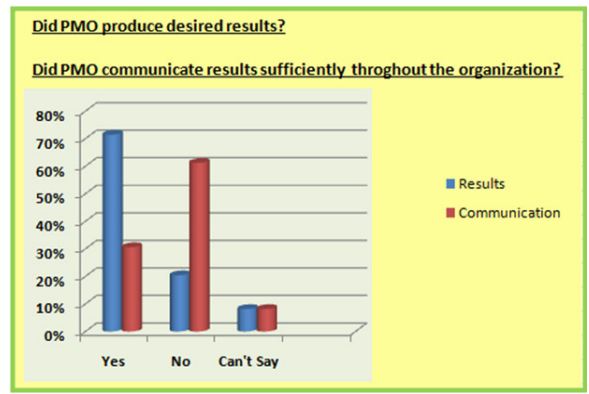

**Figure 6**: *Summary of Key Results Set #4*

### 6.1. Discussion of Key Results Area#1

- ☑ 59% of the practitioners (29 out of 49) believed that the IT PMOs they worked with had any planned project for maturity improvement of the PMO concerned.

- ☑ 18% of the practitioners (09 out of 49) confirmed that the IT PMOs they worked with did not have any planned project for maturity improvement of the PMO concerned.

- ☑ 27% of the practitioners (13 out of 49) asserted that the IT PMOs they had worked with, had on their roadmap a planned project for maturity improvement of the PMO concerned.

## 6.2. Discussion of Key Results Area#2

- ☑ 64% of the practitioners (27 out of 49) confirmed that the IT PMOs they worked with were also doing additional activities that were not aligned to originally approved PMO-charter.

- ☑ 23% of the practitioners (11 out of 49) were not sure if the IT PMOs they worked with were doing only those activities that were aligned to originally approved PMO-charter.

- ☑ 18% of the practitioners (09 out of 49) confirmed that the IT PMOs they worked with doing only those activities that were aligned to originally approved PMO-charter.

## 6.3. Discussion of Key Results Area#3

- ☑ 31% of the practitioners (15 out of 49) believed that the IT PMOs they worked with had been customized only once (initially) to align the implementation with business strategy. It was never done subsequently.

- ☑ 33% of the practitioners (16 out of 49) were not aware of any strategy that the IT PMOs they worked with, had for customization of implementation of the PMOs concerned, as a part of alignment with business, as and when business changes.

- ☑ 20% of the practitioners (10 out of 49) asserted that the IT PMOs they worked with were never required to customize their implementation of PMOs concerned, as a part of alignment with business, as and when business underwent changes.

- ☑ 16% of the practitioners (8 out of 49) asserted that the IT PMOs they worked with frequently customized their implementation of PMOs concerned, as a part of alignment with business, as and when business underwent changes. Such PMOs had a defined strategy for keeping IT PMOs in sync with business, as a rolling wave process.

## 6.4. Discussion of Key Results Area#4

- ☑ <u>Results Delivery</u>:

  - ❖ 71% of the practitioners (35 out of 49) believed that IT PMOs they worked with produced the results desired by the business.

  - ❖ 8% of the practitioners (4 out of 49) were not sure if the IT PMOs they worked with were able to produce the results desired by business or not.

  - ❖ 20% of the practitioners (10 out of 49) believed that IT PMOs they worked with did not produce the results desired by the business.

- ☑ <u>Self-Promotion</u>:

  - ❖ 61% of the practitioners (30 out of 49) believed that IT PMOs they worked with were unable to effectively communicate the results produced by them throughout the organization.

- Another 8% of the practitioners (4 out of 49) were not sure if the IT PMOs they worked with were effectively communicating the results produced by them throughout the organization.

- Only 31% of the practitioners (15 out of 49) believed that IT PMOs they worked with were effectively communicating the results produced by them throughout the organization.

## 7. Summary and Concluding Remarks

This research complements previous studies by investigating the key determinants that contribute significantly to establishing a sustainable PMO in the current scenario of global recession and cost-cuttings. study builds on our previous research on PMO[18]. Summary of our findings are as follows:

Organizations that wish to be successful in implementing a sustainable IT PMO should:

- ☑ *Continuously strive to improve the PMO Maturity Level*, following the initial implementation.

- ☑ All the activities handled by IT PMOs should be aligned to a *documented and approved PMO Charter*, enabling organization to formally acknowledge the PMO efforts and results for business value propositions.

- ☑ IT PMOs should employ and effective strategy for customization and alignments with changes in business strategy so as to remain cost effective and innovative and continue to add business value.

- ☑ IT PMOs should also focus on self-promotion within the organization by *not only* producing the results *but also* communicating the results produced professionally throughout their organization in order to earn executive support, be viewed as an organizational entity offering a competitive advantage for business and overcome skepticism.

## 8. Implications for further research

This research is a step towards further understanding the causes of failed PMO initiatives in IT-Industry and key lessons learnt to avoid these failures in fresh PMO initiatives in IT-Industry. Similar research in PMO initiatives of other industries (like construction, medicine etc.) might bring a different perspectives to the concepts acquired from this research. Research in less-successful PMOs might validate or challenge the conclusions reached in this study.

## 9. Limitations of the study

It is well known that every research/study has some limitations pertaining to methodology and dataset. In this particular case, a limitation of our study was the relatively small and convenient sample size. For this reason, our findings cannot be generalized to the broader community based on this study alone. Further empirical studies with larger data samples are needed for generalization.

## Acknowledgement

Authors would like to sincerely thank all the PMO staff and senior project managers for sparing their time and providing their valuable inputs as a part of the semi-structured interviews that were conducted and iterated to weed-out the results. Many thanks to all the peers, colleagues and organizations who supported this research by sharing their valuable time, feedback, project data and experiences on failed and challenged PMOs.

## References

[1] PMI (2008); "A guide to the project management body of knowledge", Fourth Edition. Pennsylvania: *Project Management Institute*, Inc.

[2] Martin, N.L.; Pearson, J.M.; Furumo, K.A.(2005); "IS Project Management: Size, Complexity, Practices and the Project Management Office" published in *Proceedings of the 38th Annual Hawaii International Conference on System Sciences*. HICSS '05, pp. 1 – 10, Digital Object Identifier: 10.1109/HICSS.2005.359

[3] Pellegrinelli, S., and Garagna, L. (2009) Towards a conceptualisation of PMOs as agents and subjects of change and renewal. *International Journal of Project Management*, Vol. **27** (7), 649–656.

[4] Hobbs, B., and Aubry, M. (2007) A multi-phase research program investigating project management offices (PMOs): the results of phase 1. *Project Management Journal*, Vol. **38** (1), 74-86.

[5] Aubry, M., Hobbs, B., and Thuillier, D. (2008) Organisational project management: an historical approach to the study of PMOs. *International Journal of Project Management*, Vol. **26** (1), 38–43.

[6] Hobbs, B., Aubry, M., and Thuillier, D. (2008) The project management office as an organisational innovation. *International Journal of Project Management*, Vol. **26** (5), 547–555.

[7] Hurt, M., and Thomas, J. L. (2009) Building value through sustainable project management offices. *Project Management Journal*, Vol. **40** (1), 55–72.

[8] Cooke-Davies, T. J., Crawford, L. H., and Lechler, T. G. (2009) Project management systems: moving project management from an operational to a strategic discipline. *Project Management Journal*, Vol. **40** (1), 110–123.

[9] Aubry, M., and Hobbs, B. (2011) A fresh look at the contribution of project management to organizational performance. *Project Management Journal*, Vol. **42** (1), 3–16.

[10] Thomas, J. L., and Mullaly, M. (2008a) *Researching the value of project management*. Newtown Square, PA: Project Management Institute.

[11] PM-Solutions (2012), "*The State of the PMO 2012*" Industry Research Report available on-line http://www.pmsolutions.com/research , accessed on-line.

[12] Carrillo, J.V.; Abad, M.E.; Cabrera, A.S.; Jaramillo, D.H.(2010), "Success factors for creating a PMO aligned with the objectives and organizational strategy" published in Proceedings of *IEEE International Conference*, ANDESCON,  pp. 1 – 6, Digital Object Identifier: 10.1109/ANDESCON.2010.5629937

[13] Gartner (2010),  *The Garner Industry Research* [Online]: http://www.gartner.com/technology/research.jsp.

[14] Dodson, T.; Stevenson, T.; Egebo, T.; Strykowsky, R.; Langish, S.;Williams, M. (2011); "Ensuring compliance and consistency in an R&D environment: The PPPL Office of Project Management" published in Proceedings of *IEEE/NPSS 24th International Symposium on Fusion Engineering* (SOFE), pp. 1 – 5, Digital Object Identifier: 10.1109/SOFE.2011.6052301

[15] Kendall, G., and Rollins, S.C. (2003); Advanced Project Portfolio Management and the PMO: Multiplying ROI at Warp Speed. *J. Ross Publishing*, Fort Lauderdale, Fla.
[16] Dai, C. X. (2002); "The Role of Project Management Office in Achieving Project Success", *PMI Annual Seminars and Symposium*, San Antonio, Texas, United States.
[17] Karkukly, W. (2010); "Outsourcing of PMO functions for improved organizational performance"; *Trafford publishing, Indiana,* USA.
[18] P.M. Khan, M.M.S. Beg,(2012); "Project Management Office(PMO): An Organizational Asset for Value Propositions in Software Development Projects " Proc. *International Conference on Emerging Trends in Engineering and Technology*, College of Engineering, T. M. University, Moradabad, India, E-ISBN: 978-93-8137804-5.